\documentclass[namedreferences,hyperref,optionalrh,solaromanenum]{spr-sola}

\usepackage{graphicx}                    
\usepackage{color}                       
\usepackage{ulem}

\chardef\us=`\_

\newcommand{\Alfven}{Alfv\'{e}n}

\newcommand{\V}[1]{\mathbf{#1}}

\newcommand{\ALPS}[1]{\texttt{ALPS}} 
\newcommand{\PLUME}[1]{\texttt{PLUME}}

\newcommand{\change}[1]{#1}

\begin{document}

\begin{frontmatter}

\title{Wave Emission and Absorption in a Near-Sun Proton-Cyclotron Wave Storm}

%
\author[addressref={aff1},corref,email={kgklein@arizona.edu}]
{\inits{K.G.}\fnm{Kristopher G. }\snm{Klein}\orcid{0000-0001-6038-1923}}
\author[addressref={aff3},email={}]{\inits{D.}\fnm{Daniel } \snm{Verscharen}\orcid{0000-0002-0497-1096}}
\author[addressref={aff1},email={}]{\inits{M.}\fnm{Mihailo } \snm{Martinovic}\orcid{0000-0002-7365-0472}}
\author[addressref={aff2},email={}]{\inits{N.}\fnm{Niranjana}\snm{}\orcid{0000-0002-8941-3463}}
\author[addressref={aff4}]{\inits{A.}\fnm{Ali } \snm{Rahmati}\orcid{0000-0003-0519-6498}}
\author[addressref={aff4}]{\inits{R.}\fnm{Roberto } \snm{Livi}\orcid{0000-0002-0396-0547}}
\author[addressref={aff4}]{\inits{D.}\fnm{Davin } \snm{Larson}\orcid{0000-0001-5030-6030}}
\author[addressref={aff5}]{\inits{M.}\fnm{Michael } \snm{Stevens}\orcid{0000-0002-7728-0085}}


\runningauthor{K.G.~Klein and others}
\runningtitle{\textit{Solar Physics} A Near-Sun Proton Cyclotron Wave Storm}

\address[id=aff1]{Lunar and Planetary Laboratory, University of Arizona, Tucson, AZ, USA}
\address[id=aff3]{Mullard Space Science Laboratory, University College London, Dorking, UK}
\address[id=aff2]{Department of Physics, University of Arizona, Tucson, AZ, USA}
\address[id=aff4]{Space Science Laboratory, University of California, Berkeley, CA, USA}
\address[id=aff5]{Smithsonian Astrophysical Observatory, Cambridge, MA, USA}

\begin{abstract}
Quantification of energy transport and dissipation in weakly collisional heliospheric plasmas that are far from local thermodynamic equilibrium is an outstanding scientific problem.
A central challenge is determining how non-Maxwellian
velocity-space structure affects damping and emission of coherent ion-scale waves, especially compared to simplified analytical models for background plasma velocity distributions.
In this work, we study the damping and emission of parallel-propagating proton cyclotron waves for two models of proton velocity distributions measured by the SPAN-I instrument on board Parker Solar Probe during an extended storm of waves with left-hand polarization in the solar wind at a heliocentric distance of 30.1 solar radii.
Using the measured velocity distribution rather than a two-component bi-Maxwellian model predicts instabilities consistent with the observed coherent waves.
For intervals in which both models predict net damping, the observed VDF model yields weaker damping in 90\% of cases, with a reduction in the integrated heating rate of 0.44 relative to the bi-Maxwellian model. 
These results suggest that simplified analytical velocity distribution models may overestimate cyclotron damping and underestimate wave emission in the near-Sun solar wind.
\end{abstract}

%

\end{frontmatter}

%
\section{Introduction}
\label{sec:intro} 
Collision-poor plasmas, such as the solar wind, frequently exhibit non-Maxwellian velocity distribution functions \citep[VDFs;][]{Marsch:2012,Verscharen:2019}.
Temperature anisotropies, secondary populations, and more general departures from local thermodynamic equilibrium are interpreted as signatures of dissipation processes \citep{He:2015,Howes:2022} and serve as free energy sources to drive micro-instabilities \citep{Chen:2016,Klein:2018,Martinovic:2021b,Niranjana:2026}.
Determining the relation between these structures and the emission of ubiquitously observed coherent waves and extended wave storms \citep[e.g.][]{Jian:2009,Wicks:2016,Gary:2016} is necessary to  characterize the behavior of the solar wind, particularly in the inner heliosphere \citep{Klein:2021,Bowen:2022,McManus:2024}
and when the structures are not well described by analytical functions \citep{Walters:2023,Klein:2026,Ran:2026}.

In addition to characterizing their generation, coherent waves have been proposed to play a significant role in the dissipation of electromagnetic energy and the heating of the expanding solar wind.
One of the main contenders for this dissipation is resonant cyclotron heating, which has been observed in selected intervals measured by Parker \citep{Bowen:2020b,Bowen:2022,Bowen:2024a,Bowen:2024b}
and has been inferred to provide sufficient energy to account for the observed radial temperature trend remaining shallower than expected for adiabatic expansion alone \citep{Niranjana:2024}.
However, calculations of cyclotron heating rates typically assume a bi-Maxwellian VDF, which is not consistent with the observed VDFs in the solar wind.
Wave-particle interactions deform the VDF from a Maxwellian.
This deformation impacts how efficiently a wave is able to damp \citep{Isenberg:2012,Ran:2026}, making the inclusion of non-Maxwellian features in calculations of wave behavior essential.

For decades, determining the linear plasma response relied on assuming a particular analytical form for the background VDF, e.g. bi-Maxwellian \citep{Roennmark:1982,Verscharen:2018a,Klein:2025-PLUME} or kappa \citep{Astfalk:2015,Lopez:2021} functions.
Recently, a variety of numerical tools that do not assume a particular analytical form have been developed to assess the impact of non-Maxwellian structure in the VDF on wave emission and damping \citep[e.g.][]{Astfalk:2017,Xie:2019}.
In this work, we use the Arbitrary Linear Plasma Solver \citep[\texttt{ALPS};][]{Verscharen:2018} to characterize wave frequencies and damping rates for proton VDFs measured by the SPAN-I instrument on Parker Solar Probe (Parker) at a time interval coincident with coherent left-hand polarized ion-scale waves measured by the FIELDS instrument.

A brief overview of \ALPS\ \ and SPAN-I measurement processing is provided in \S~\ref{sec:methods}.
The analysis of an example 20-minute ion-scale wave storm is presented in \S~\ref{sec:results}, with concluding discussions in \S~\ref{sec:discussion}.

\section{Methodology}
\label{sec:methods} 

\subsection{The Arbitrary Linear Plasma Solver (\texttt{ALPS})}
\label{ssec:methods.alps} 

The Arbitrary Linear Plasma Solver (\texttt{ALPS}, \cite{Verscharen:2018}) is a parallelized numerical code that solves the linear Vlasov-Maxwell dispersion relation in hot (even relativistic) magnetized plasma.  
\texttt{ALPS} allows for any number of particle species with arbitrary gyrotropic background VDFs supporting waves with any direction of propagation with respect to the background magnetic field. 
The wave equation is solved by numerically integrating the velocity gradients of the input VDFs $f_j(v_\perp,v_\parallel)$, and then identifying the frequencies $\omega(k_\perp,k_\parallel)$ that satisfy the dispersion relation. 
\texttt{ALPS} also determines the power absorbed or emitted by each plasma species with its background VDF $f_j$ and the eigenfluctuations of density $\delta n_j$, velocity $\delta \V{U}_j$, and electric and magnetic fields $\delta \V{E}$ and $\delta \V{B}$. 
The code has been updated to allow the use of Chebyshev polynomials to  represent the analytic continuation necessary for calculating the behavior of moderately damped solutions more accurately \citep{Klein:2025-ALPS}.
\texttt{ALPS} has been used to characterize wave behavior from non-Maxwellian VDFs in kinetic simulations \citep{Bianco:2026,Zhang:2025,Martinovic:2026}, from analytical functions \citep{Schroder:2025,Tischmann:2026}, and a variety of spacecraft observations, including 
Wind \citep{Walters:2023},
MMS \citep{Jiang:2024,Afshari:2024}, and
Solar Orbiter \citep{Ran:2026}.
Previous applications of \texttt{ALPS} to Parker/SPAN-I \citep{Klein:2026} focus on selected VDFs with different features and types of waves to establish the methodology for processing the measurements and interpreting the results.
In this work, we focus on an application to a larger volume of measurements focused on a single wave type: parallel-propagating proton cyclotron waves (PCWs).
This approach allows us to quantify differences between waves represented by bi-Maxwellian and observed VDF models statistically, and to compare the predictions to the observed electromagnetic fields.

\subsection{\change{Wave Storm Identification}}
\label{ssec:methods.waves} 

Parker \citep{Fox:2015} was launched in 2018 to characterize plasma processes in the inner heliosphere as the solar wind is heated and accelerated into the solar system.
For this study, we use magnetic field measurements from the FIELDS fluxgate magnetometer \citep{Bale:2016} and proton velocity distributions from the SWEAP/SPAN-I electrostatic analyzer \citep{Livi:2022}.
\change{Electron density, used to calibrate the inertia term in the \Alfven \ velocity, is derived from quasi-thermal noise (QTN) spectra \citep{Moncuquet:2020}.}

\begin{figure}
    \centering
    \includegraphics[width=0.95\linewidth]{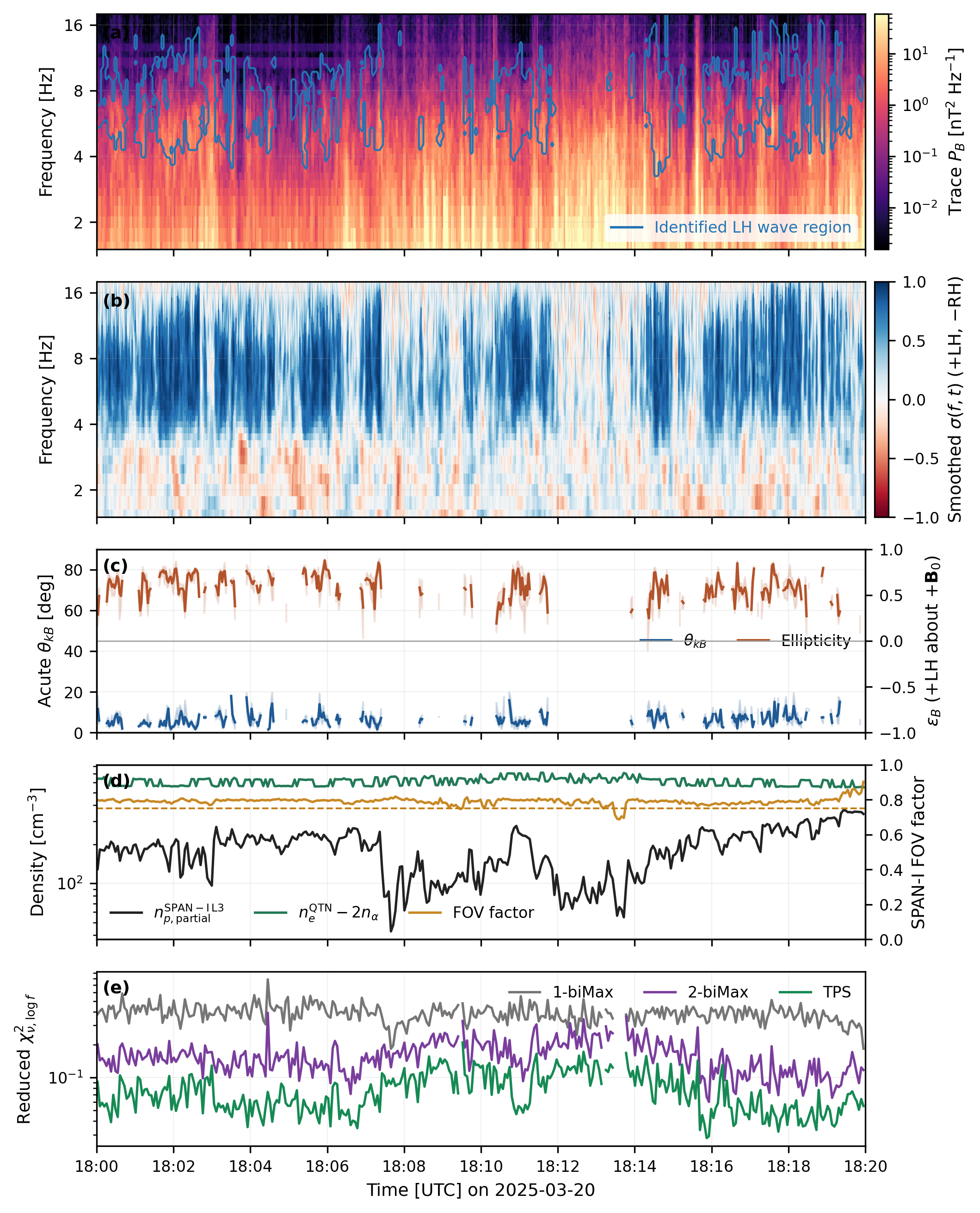}
    \caption{\change{Summary of ion-cyclotron wave storm observations.
    (a) Trace Magnetic Field Power Spectral Density. 
    Identified regions of coherent left-handed waves are outlined in blue.
    (b) Smoothed magnetic field spacecraft-frame polarization.
    (c) Power-weighted median presumed acute wave-normal angle $\theta_{kB}$
     and magnetic ellipticity 
     $\epsilon_B$
     within the identified left-handed wave regions; shaded bands show the corresponding interquartile ranges.
    (d) SPAN-I and QTN proton densities and SPAN-I FOV factor.
     (e) Goodness-of-fit for one- and two-component bi-Maxwellian and thin plate spline (TPS) models, Eqn.~\ref{eqn:chi2}.
    }
    }
    \label{fig:observations}
\end{figure}

We focus on a twenty-minute interval of SPAN-I observations on 2025-03-20, 18:00:00-18:20:00, when Parker was at 30.1 Solar Radii ($R_\odot$, 0.14\,au).
\change{An overview of this interval is shown in Fig.~\ref{fig:observations}, with panel (a) showing the trace magnetic field power spectral density.}
Applying a wavelet transform to 
\change{calculate the coherence of the magnetic field fluctuations in the $X-Y$ plane transverse to the background magnetic field $\sigma$}
\citep{Torrence:1998,Niranjana:2024} reveals a consistent and coherent wave storm at the proton-cyclotron frequency with left-handed polarization throughout much of this time interval; 
\change{see blue regions in Fig.~\ref{fig:observations}(b).
To further characterize these coherent waves, we use singular-value decomposition (SVD) to estimate the wave-normal direction \citep{Santolik:2003} and to calculate the acute angle $\theta_{kB}$ between the wavevector and the local background magnetic field $B_0$.
The magnetic ellipticity $\epsilon_B$ is calculated from the relative amplitudes and phase difference of the two transverse wavelet components, with positive values corresponding to left-handed polarization about $B_0$. 
For each time, the plotted curves and shaded regions show the power-weighted median and inter-quartile range for $\theta_{kB}$ and $\epsilon_B$ across the frequencies lying within the identified wave regions. 
Gaps occur when no time-frequency pixels satisfy the wave-selection criteria of $|\sigma|>0.7$;
more details on wave-selection are found in \cite{Niranjana:2026}.
The generally small $\theta_{kB}$
 values and positive ellipticities approaching unity indicate that the observed fluctuations are predominantly quasi-parallel and nearly circularly left-hand polarized. 
}

\subsection{\change{Measurement Processing}}
\label{ssec:methods.spani} 

\change{
To ensure SPAN-I has a consistently good field of view (FOV) during this interval, we implement the methodology described in detail by \citet{Romeo:2024} and \citet{Yogesh:2026}.
To calculate a FOV goodness metric, we fit the summed differential energy flux ($J_E$) as a function of each angle to a one-dimensional Gaussian function and then compute the integrated area of the product of the fitted Gaussians within the angular limits of the instrument.
If this integrated area is near unity, most of the proton VDF is within the instrument's FOV.
As shown in panel (d) of Fig.~\ref{fig:observations}, nearly all of our intervals have an FOV factor of $>0.75$.
}

\texttt{ALPS} requires a uniform distribution of phase-space density on a gyrotropic grid, $(v_\perp,v_\parallel)/v_A$,
where $v_A= B/\sqrt{4 \pi n_p m_p}$ is the proton \Alfven \ velocity.
In order to produce this grid, for each of the 342 SPAN-I  measurements in the L2 dataset from 18:00:00-18:20:00, we first transform the differential energy flux $dQ/(dE \,d\Omega)$ into a phase-space density $f_p(\V{v})$.
To reduce statistical noise, we mask out any points for which the differential energy flux is associated with a single particle count.
For observations from later encounters, which are closer to the Sun, the proton flux is high enough that this one-count restriction does not limit our ability to resolve the proton VDF out to several thermal widths.
We then calculate the solar wind bulk velocity as the first velocity moment of $f_p(\V{v})$ and shift the distribution into the plasma frame.
$f_p(\V{v})$ is then rotated into a local magnetic field aligned co\"ordinate system, and the two transverse directions are gyrotropized as $v_\perp = \sqrt{v_{\perp,1}^2+v_{\perp,2}^2}$.

For our comparison model (BiMax), this field aligned, gyrotropized proton VDF is fit with a two-component, core-and-beam, bi-Maxwellian model.
As the electron density is measured more consistently using quasi-thermal noise spectroscopy than the ion densities from SPAN-I, for the inertial term in the $v_A$, we define the proton density as
\begin{equation}
    n_p=n_e^{QTN}-2 n_\alpha
\end{equation}
where $n_\alpha$ is the SPAN-I measured density for fully-ionized Helium.
\change{As shown in panel (d) of Fig.~\ref{fig:observations}, using} the SPAN-I partial proton density moments leads to significant variations in $v_A$, impacting the widths of the normalized VDFs.

 \begin{figure} 
 \centerline{\includegraphics[width=1.0\textwidth,clip=]{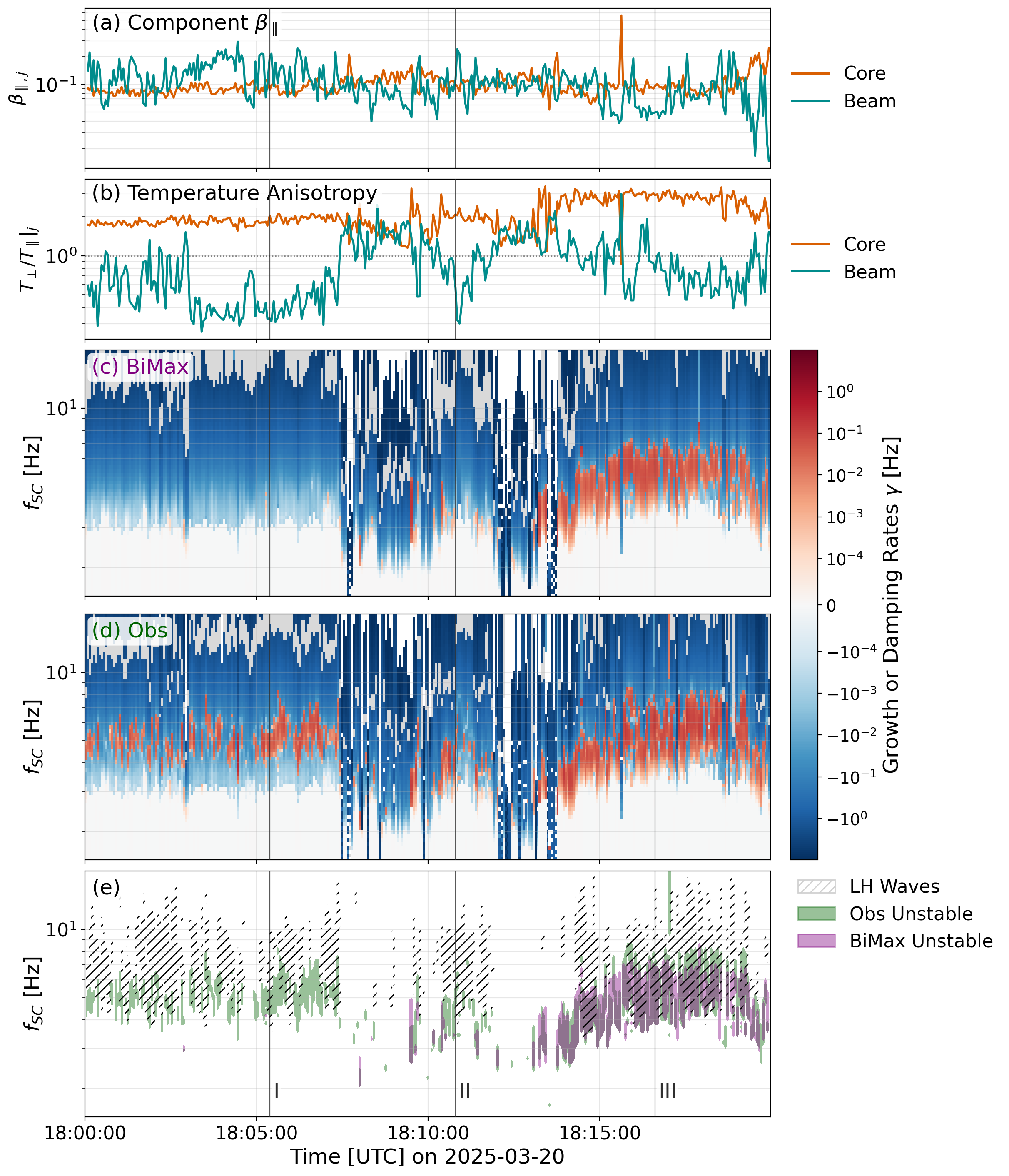}}
 \caption{
Fit parameters and growth and damping rates for Alfv\'en/proton cyclotron waves that propagate parallel to the magnetic field and into the anti-Sunward direction.
\change{(a) Plasma $\beta_{\parallel,j}$ and (b) Temperature anisotropy $T_{\perp,j}/T_{\parallel,j}$ for proton core and beam bi-Maxwellian fits.}
(c) Growth (red) and damping (blue) rates as calculated for the two-component bi-Maxwellian model.
(d) Growth and damping rates for the observed VDF from \texttt{ALPS}. 
\change{Grey regions indicate overdamped regions when $|\gamma/\omega_r|>0.1$.}
(e) Overlay of the observed coherent left-handed power frequency distribution (cross-hatched, \change{see Fig.~\ref{fig:observations}a and b}) and regions of instability with  $\gamma^{\textrm{Obs}}$ (green) and $\gamma^{\textrm{Bi-Max}}$ (purple) greater than $10^{-3}$ Hz.
 }
 \label{fig:spectra}
 \end{figure}

We restrict our fits to have $T_{\perp,j}/T_{\parallel,j} \in [0.1,10]$ and $T_{\parallel,c}/T_{\parallel,b} \in [0.1,10]$, but place no restrictions on the relative densities or inter-component drift speeds.
The fit values for the component $\beta_{\parallel,j}=8 \pi n_j T_{\parallel,j}/B^2$ and temperature anisotropy $T_{\perp,j}/T_{\parallel,j}$ are shown in Fig.~\ref{fig:spectra}, panels a and b.
We also consider a single-component bi-Maxwellian model, but the simpler model for these intervals always produces a worse fit compared to the two-component model when evaluated with 
\begin{equation}
\chi_{\nu,\log f}^{2}
=
\frac{1}{N-k}
\sum_{i=1}^{N}
\left[
\log_{10} f_{p,i}^{\mathrm{meas}}
-
\log_{10} f_{p,i}^{\mathrm{model}}
\right]^2 .
\label{eqn:chi2}
\end{equation}
\change{The values of $\chi^2_{\nu,\log f}$ for both the one- and two-component models are shown in Fig.~\ref{fig:observations}(e).
The number of measured points is $N$, and the number of model parameters is $k=4$ and $k=8$ for the one- and two-component fits.}

To ensure a smooth continuation of the VDF to large velocities, we evaluate the phase-space density from the two-component model along velocity rings beyond the measured velocity points.
A detailed explanation of this processing, \change{including the impact of velocity resolution and smoothing of the TPS,} is provided in the supplemental information appendix of \cite{Klein:2026}.
The measured points and collars for selected intervals are shown in panels e and f of Figs.~\ref{fig:int1}, \ref{fig:int2}, and \ref{fig:int3}.
The width and number of rings in the collar do not qualitatively impact the resulting dispersion relations.
Lastly, a thin-plate spline interpolation (TPS) is applied to the measured points and collar, creating a Cartesian grid of $f_j$ in velocity space using the mid-point of the collar as the minimum and maximum of the velocity range. 
\change{
For the TPS, we follow \cite{Hutchinson:1989} and estimate the effective number of fitted parameters as $k_{\mathrm{TPS}}=3+\mathrm{tr}[K(K+\lambda I)^{-1}]$, where $K$ is the thin-plate-spline kernel matrix evaluated at the measured VDF points and $\lambda=0.05$ is the smoothing parameter.
The trace measures the spline’s effective flexibility, while the additional three degrees of freedom represent its affine polynomial component.
Fig.~\ref{fig:observations}(e) shows that $\chi^2_{\nu,\log f}$ for the TPS model is consistently smaller than either of the bi-Maxwellian fits.
}
For this work, \change{we use} a grid of 401 points in $v_\parallel/v_A$ and 200 points in $v_\perp/v_A$.

For this work, we focus on the weakly damped, parallel-propagating solutions supported by the measured VDF and the equivalent bi-Maxwellian model, giving us fast magnetosonic/whistler and Alfv\'en/proton cyclotron (PCW) waves that propagate in the Sunward and anti-Sunward directions.
The anti-Sunward PCW is the only parallel-propagating solution that matches the observed frequencies and the observed left-handed polarization of the wave storm when the dispersion relations are Doppler-shifted from the plasma to the spacecraft frame.
\change{The Appendix illustrates the lack of matching for the other three parallel wave solutions for three selected intervals.}
We therefore focus on the behavior of this solution alone for the two models in the remainder of the paper.

\section{Results}
\label{sec:results} 

Panel c of Fig.~\ref{fig:spectra} shows the imaginary part $\gamma$ of the frequency for the anti-Sunward PCW as a function of time and spacecraft-frame frequency from the bi-Maxwellian model. 
Positive and negative values of $\gamma$ represent unstable and damped waves respectively.
The bi-Maxwellian solutions are entirely stable for the first nine minutes of the wave storm.
The core temperature anisotropy is around two for most of this interval, and with $\beta_{\parallel,c}\sim 0.1$, making the core stable compared to the $\gamma=10^{-3}\Omega_p$ PCW threshold from \cite{Verscharen:2016}, where $\Omega_p = qB/m_p c$ is the proton cyclotron frequency.
The anisotropy increases with intermittent spikes over the next several minutes, at which point it stabilizes around $3$.
This anisotropy is large enough to drive the PCW unstable, with growth rates of order $10^{-3}-10^{-2}$\,Hz.

Panel d of Fig.~\ref{fig:spectra} shows $\gamma$ for the anti-Sunward PCW from the observed VDF model.
This model predicts unstable PCWs for times spanning the entire twenty minutes considered, including the early times that the bi-Maxwellian model considers stable.
After 18:12:00, both models have similar unstable frequency-time regions, though the observed model has moderately higher growth rates.

Panel e of Fig.~\ref{fig:spectra} compares the unstable regions of moderate growth, selected as $\gamma>10^{-3}$ Hz, from the bi-Maxwellian and observed models, purple and green contours respectively, against the observed left-handed wave power, shown as grey cross-hatching.
The coherent power calculations follow Appendix B of \cite{Niranjana:2023}, where a wavelet transform of the FIELDS magnetometer data is performed.
The vector wavelet components are rotated into field aligned co\"ordinates, and the correlation between the phases of the two transverse components is evaluated.
If the correlation is greater than $0.7$ for a given frequency-time region, that area is identified as having coherent left-handed wave power.
We see consistent left-handed wave power through the entire 20-minute interval, typically ranging from 5 to 12 Hz.
There is a disappearance of the observed power for short intervals around 18:07 and 18:12.

The observed model predicts growing PCWs for nearly all of the intervals when left-handed wave power is measured.
The Doppler-shifted frequencies of the predicted unstable waves align well with the lower-frequency extent of the observed coherent waves.
As noted above, the bi-Maxwellian model does not predict unstable waves for the first half of the twenty-minute interval.
Both models make similar predictions for the occurrence and frequency extent of unstable PCWs in the latter half of the interval, though the observed-model peak growth rates are enhanced.

\subsection{Selected Intervals}
\label{ssec:intervals} 

We select three SPAN-I measurement intervals, highlighted in Fig.~\ref{fig:spectra}; 
I: 18:05:23, 
II: 18:10:48, and 
III: 18:16:37, 
to examine in detail.

\begin{figure}
    \centering
    \includegraphics[width=1.0\linewidth]{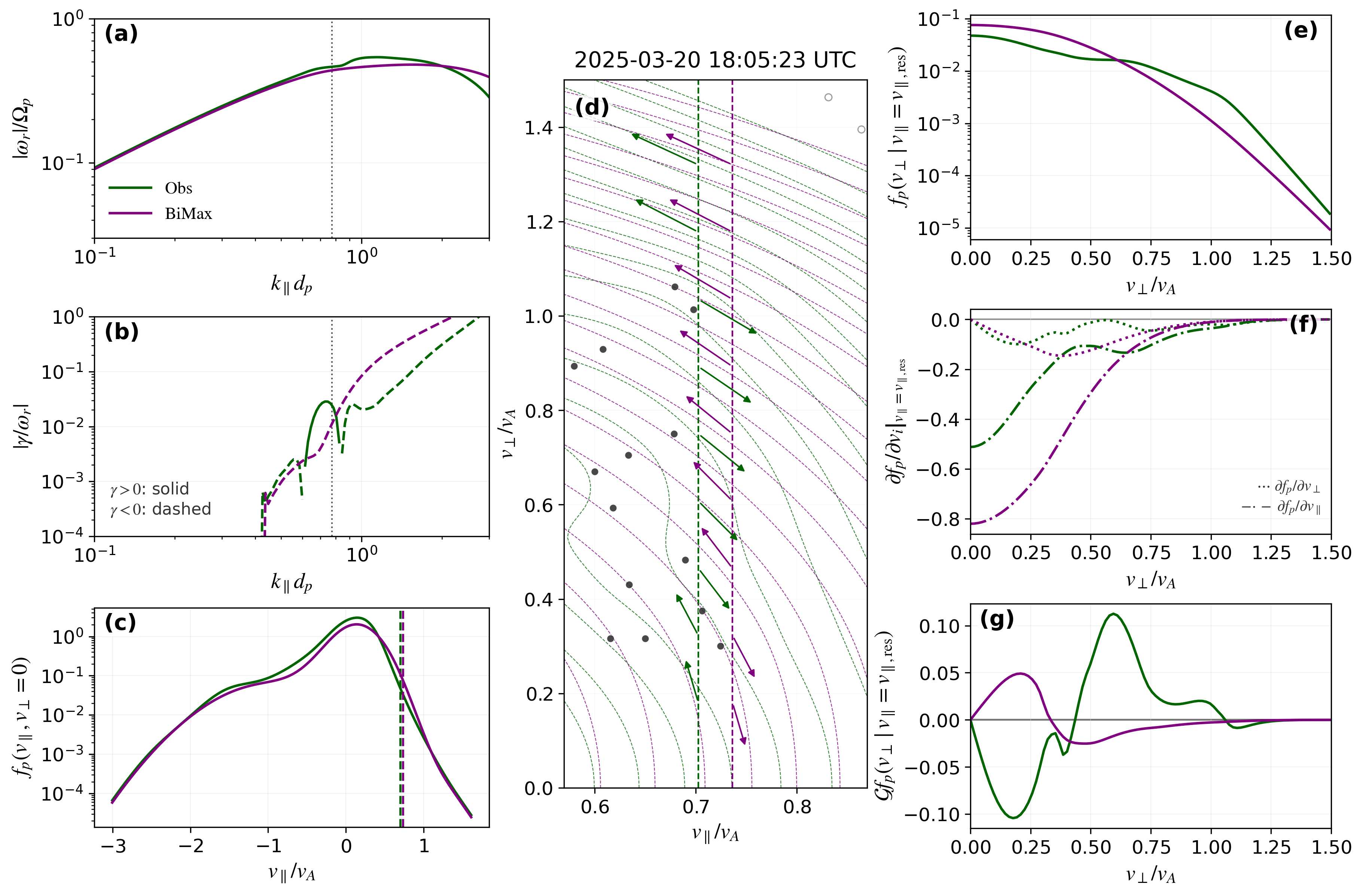}
    \caption{VDF characteristics and anti-Sunward PCW dispersion relation associated with bi-Maxwellian (purple) and observed (green) models for Interval I:
    a) real frequency $|\omega_r|/\Omega_p$,
    b) growth ($\gamma>0$) and damping ($\gamma<0$),
    c) parallel slice $f_p(v_\perp=0,v_\parallel)$,
    e) perpendicular slice $f_p(v_\perp,v_\parallel=v_{\textrm{res}})$,
    f) velocity gradients $\partial{v_\perp}f(v_\perp)$ and $\partial_{v_\parallel}f(v_\perp)$ evaluated at $v_\parallel = v_{\rm res}$,
    g) quasilinear operator $\mathcal{G}(v_\perp)$ applied to $f_j$, Eqn.~\ref{eqn:G} applied at the same velocities as panel f,
    d) VDF models Obs (green) and BiMax (purple), shown as contours, along with arrows indicating particle diffusion along tangents of semi-circles centered at $\omega_r/k_\parallel$.
    }
    \label{fig:seven}
\end{figure}

\begin{figure} 
 \centerline{\includegraphics[width=1.0\textwidth,clip=]{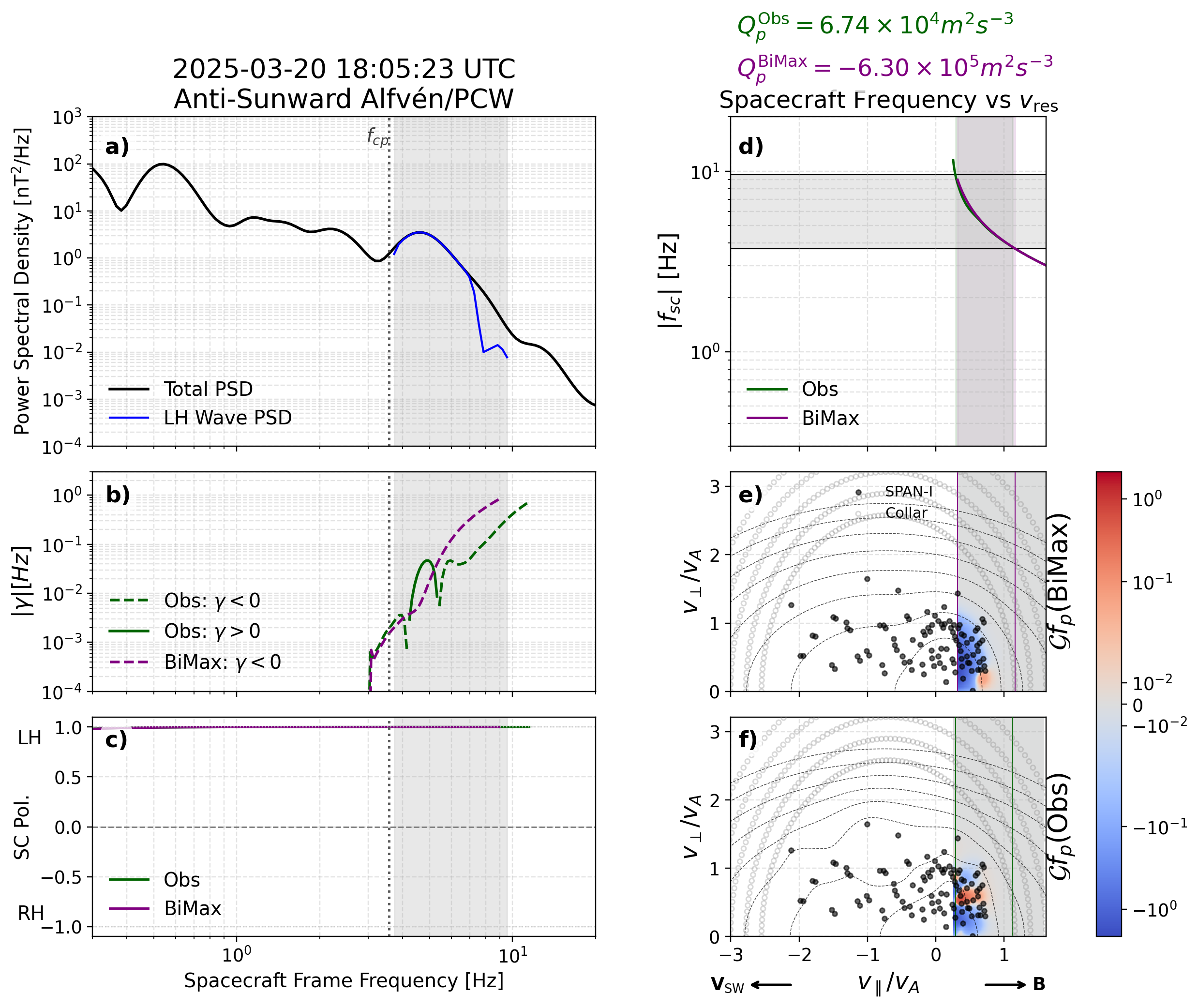}}
 \caption{
Comparison of proton VDF, coherent wave power, and predicted linear growth and damping for the two VDF models on 2025-03-20 at 18:05:23.
a) Total power spectral density (black) and coherent left-handed power \change{(blue)}.
b) Calculated growth (solid lines) and damping (dashed) rates as functions of spacecraft frame frequency for the Obs (green) and two-component biMax (purple) models.
c) Spacecraft frame polarization.
d) Spacecraft frame frequency as a function of resonant velocity.
e) Two-component biMax model VDF with quasilinear emission and absorption $\mathcal{G}f_j$.
f) Same as panel e, for observed SPAN-I VDF model.
In panels a-c, vertical grey regions indicate frequency extent of left-handed wave power. 
In panels d-f, vertical lines indicate overlap between observed coherent waves and resolved resonant velocities, 
with grey shading indicating overlap with the linear solution.
 }
 \label{fig:int1}
 \end{figure}

Interval I, 18:05:23, is shown in Figs.~\ref{fig:seven} and ~\ref{fig:int1}.
Panels a and b of Fig.~\ref{fig:seven} show the real frequency $\omega_{\rm r}/\Omega_p$ and the ratio between the damping and growth rates $\gamma/\omega_{\rm r}$ for both models.
The real frequencies for the two models are qualitatively similar, with minor deviations above $k_\parallel d_p \sim 1$, where $d_p$ is the proton inertial length.
The bi-Maxwellian model is damped for all investigated wavevectors, while the observed model is strongly unstable near $k_\parallel d_p \sim 0.7$.
Panel c shows a cut through the two VDF models at $v_\perp=0$ as a function of $v_\parallel/v_A$.
Panel e shows a cut through the two VDF models at 
the normalized resonant velocity
\begin{equation}
    \frac{v_{\parallel,\rm{res}}}{v_A}=\frac{\omega_r/\Omega_p-n}{k_\parallel d_p},
\end{equation}
for $k_\parallel d_p=0.775$ as a function of $v_\perp/v_A$.
The selected wavevector and resonant velocities are shown in panels a-c.
Panel d shows isocontours of both models as a function of $(v_\perp,v_\parallel)$ near the resonant velocities.

To identify the VDF structures responsible for the different behavior, we evaluate the quasilinear operator \citep{Kennel:1966,Verscharen:2013c}
\begin{equation}
\mathcal{G}
\equiv
\left(
1 - \frac{k_\parallel v_\parallel}{\omega_r}
\right)
\frac{\partial}{\partial v_\perp}
+
\frac{k_\parallel v_\perp}{\omega_r}
\frac{\partial}{\partial v_\parallel}
\label{eqn:G}
\end{equation}
applied to the observed and bi-Maxwellian VDFs.
Negative values of $\mathcal{G}f_j$ indicate energy transfer from the resonant wave to that particular $(v_\perp,v_\parallel)$ VDF region, and positive values indicate energy transfer from the VDF to the wave.
For the cyclotron-resonant process modeled here, at a given $v_{\parallel, \rm res}$, particles are scattered locally tangent to semi‐circles in velocity space with origins centered at the parallel phase velocity of the wave, $(v_\perp, v_\parallel) = (0, \omega_{\rm r}/k_\parallel)$.
These tangent arrows are shown in Fig.~\ref{fig:seven}, panel d for both models for the selected $k_\parallel d_p$.
If the particles are scattered to higher (lower) kinetic energy, the wave loses (gains) energy. 
Integrating over $v_\perp$ yields the net energy emission or absorption for the resonant wave.
In Fig.~\ref{fig:seven}, panels f and g, we see $\mathcal{G}f_p(v_\perp)$ has distinct structure for the two models, with low $v_\perp$ particles emitting (absorbing) energy for the bi-Max (Obs) models, while absorbing (emitting) energy for $v_\perp>0.5 v_A$.
Integration over all $v_\perp$ produces the signature seen in panel b: a stable PCW for the bi-Maxwellian and an unstable PCW for the observed VDF.

We next consider which of the two model predictions are more consistent with the observed coherent waves.
The magnetic field power spectral density (PSD) for Interval I is shown in panel a of Fig.~\ref{fig:int1} as a black line, with the left-handed contribution shown in \change{blue}.
The left-handed coherent power creates a peak in the PSD and spans a limited frequency range, typical of these wave storms.
The predicted wave growth or damping as a function of spacecraft frequency is shown in panel b for the two models.
For these plots, we limit the dispersion relation to wavevectors with $|\gamma|\leq 0.5|\omega_r|$, i.e., to times at which the solution is not evanescent.
The biMax model is damped over the entire frequency range where the left-handed waves are observed, indicated as a grey vertical region, while the observed VDF model has a narrow region of instability with its peak at a frequency near the maximum of the observed left-handed wave power.
For completeness, we show the polarization of the wave solutions in panel c; both solutions are left-handed in the spacecraft frame.

Panel d shows the spacecraft frame frequency of the dispersion relations as a function of $v_{\parallel,\rm res}$.
For the entire twenty-minute interval, the magnetic field is pointed Sunward, so the anti-Sunward PCWs
propagate backwards with respect to $\V{B}$.
We therefore choose $n=1$ for calculating the resonant velocity.
We select the resonant velocities associated with the observed left-handed power, shown as vertical bands in panel d and vertical colored lines in panels e and f.

The points in gyrotropic velocity space where SPAN-I measures a differential energy flux are shown as solid points in panels e and f, with the fitted rings collaring the observed distribution shown as open circles.
Contours from the thin-plate spline interpolation are shown for the biMax and observed model in panel e and f respectively.
The velocities at which the observed wave power is resonant with the VDF are bracketed by colored vertical lines.

We plot in color in panels e and f the value of $\mathcal{G}f_p$ from Eqn.~\ref{eqn:G}.
For the biMax VDF, $\mathcal{G}f_p$ varies primarily with $v_\parallel$ and its $v_\perp$-integrated value is negative over the entire resonant velocity range.
The observed VDF model has a much more complex structure, mixing regions of emission and absorption, including a patch of emission near $v_\parallel/v_A \sim v_\perp /v_A \sim 0.75$, which is responsible for the instability shown in panel b.

To determine the net energy absorption or emission rate $Q_p^{\rm model}$, we calculate
\begin{equation}
    Q_p^{\rm model}=\int df_{sc} {\rm PSD}^{\rm LH}(f_{sc}) \gamma_p^{\rm model}(f_{sc}),
    \label{eqn:Q}
\end{equation}
where PSD$^{\rm LH}$ is the spectral density of the coherent left-handed power and $f_{sc}$ is the spacecraft frame frequency.
Even when unstable waves are predicted for a model, $Q_p$ is frequently negative given the broader frequencies that the observed left-handed waves span.
The quantity $Q_p$  gives us a clean method for quantitatively comparing the expected coupling between the left-handed waves and the model VDFs.
For this first interval,  the observed model predicts net emission of energy from the VDF and the biMax model predicts net absorption. 

\begin{figure} 
 \centerline{\includegraphics[width=1.0\textwidth,clip=]{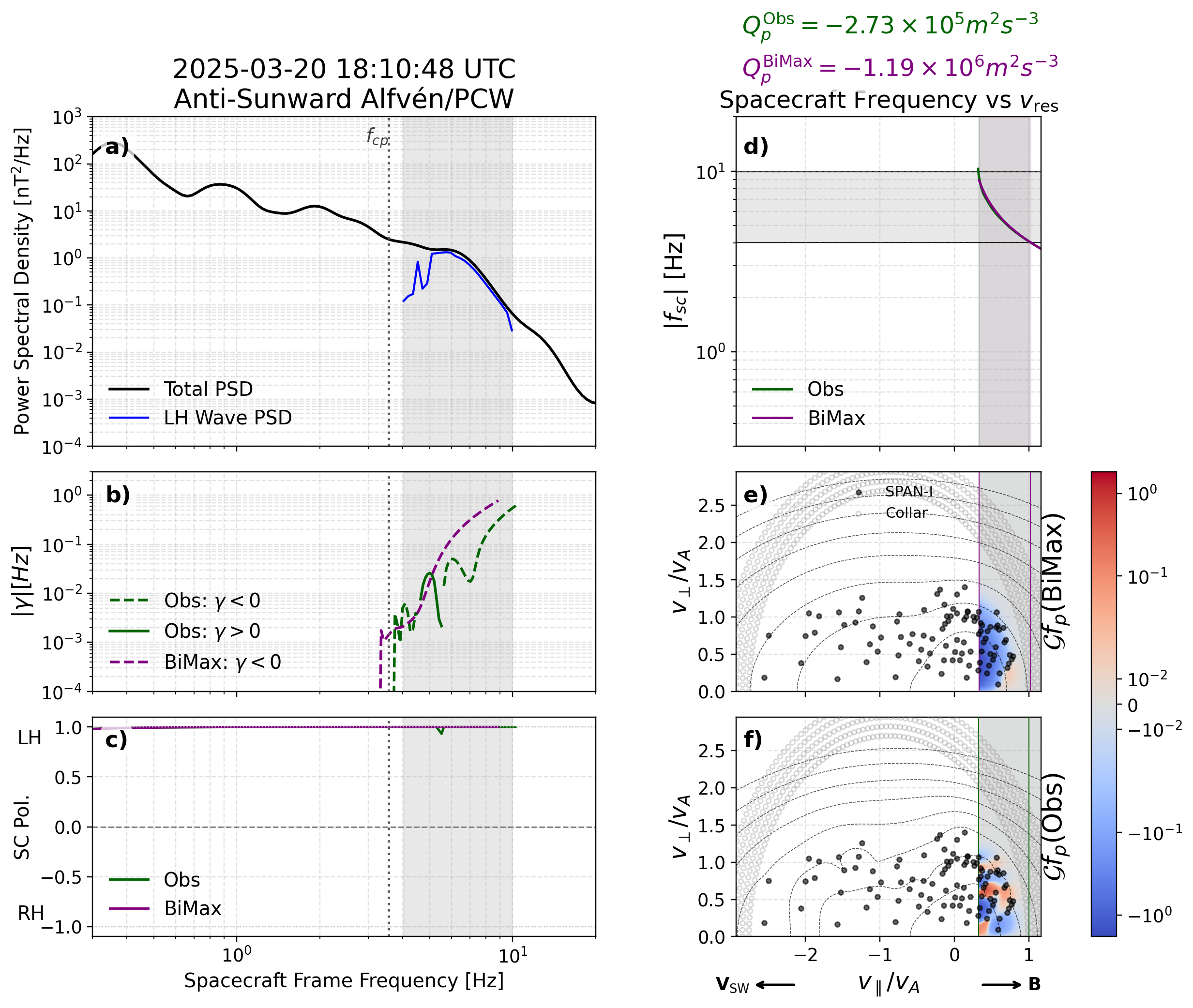}}
 \caption{
Same as Fig.~\ref{fig:int1}, but for 2025-03-20 18:10:48.
 }
 \label{fig:int2}
\end{figure}

We next consider another case, Interval II at 18:10:48, shown in Fig.~\ref{fig:int2}, following the same format as the previous figure.
This interval has clear left-handed coherent wave power, predictions from the biMax model for pure damping and from the Obs model for a narrow range of unstable wavevectors, associated with complex, localized structure in the observed VDF.
Unlike interval I, however, the integrated energy transfer rate $Q_p$ is negative for both models.
The damping is relatively weaker for the Obs model, due to a combination of the unstable solutions and weaker damping rates for the solutions with $\gamma<0$.

  \begin{figure} 
 \centerline{\includegraphics[width=1.0\textwidth,clip=]{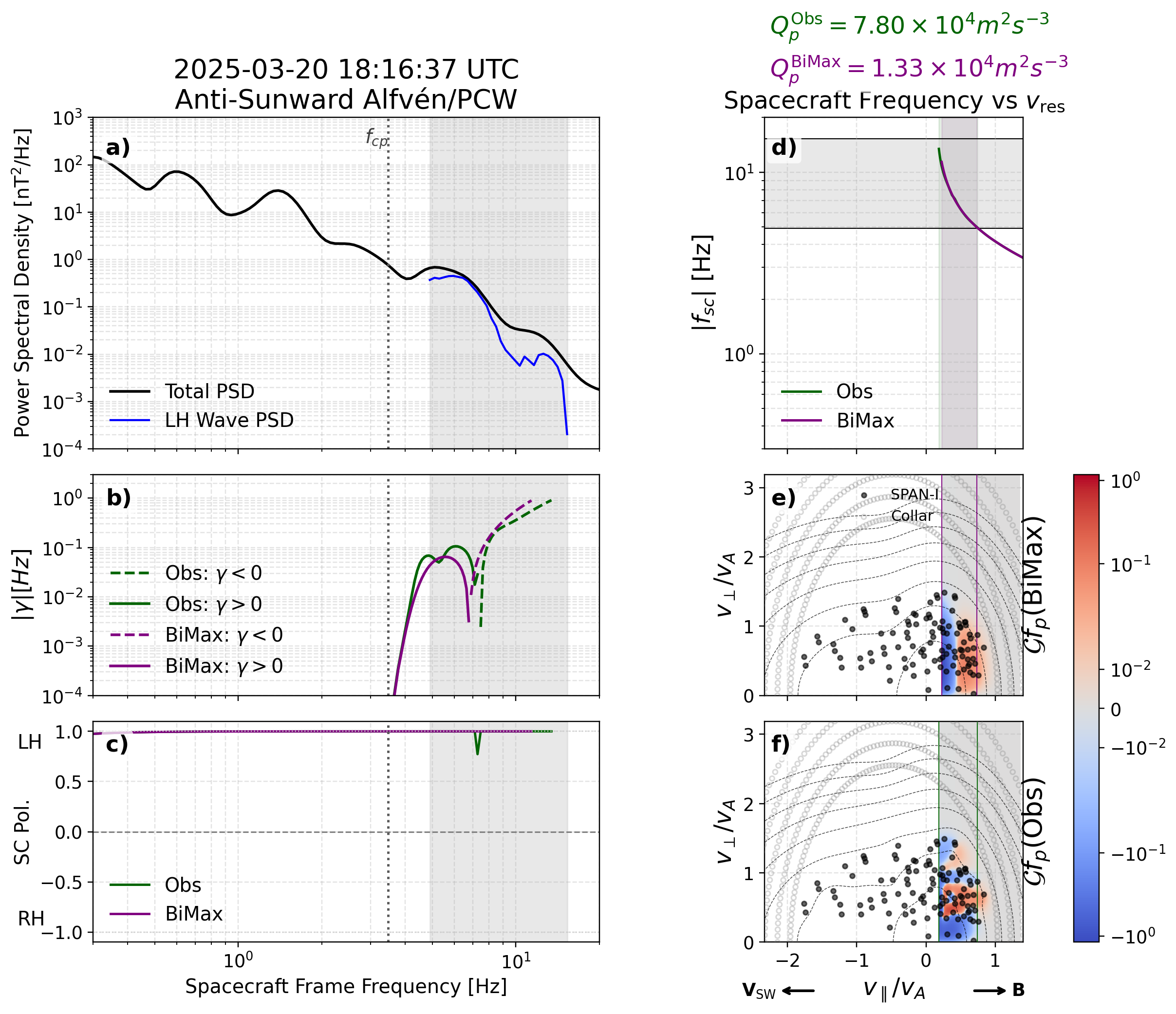}}
 \caption{
Same as Fig.~\ref{fig:int1} and \ref{fig:int2}, but for 2025-03-20 18:16:37.
 }
 \label{fig:int3}
 \end{figure}

Interval III, 18:16:37, shown in Fig.~\ref{fig:int3} following the same format as the previous two figures, again has left-handed coherent wave power present.
Unlike the two previous intervals, for this case, both models have unstable solutions and predict $Q_p>0$.
The VDF structure responsible for the unstable modes is different for the two models.
For the biMax model, the VDF emits or absorbs power for all $v_\perp$ values for a selected $v_\parallel$, changing over from unstable to stable at $v_\parallel \sim 0.5 v_A$.
For the Obs model, we again have a horizontal band near $v_\perp \in (0.75,1.0) v_A$ that emits power, with bands of absorption below and above.
The Obs model predicts a slightly more unstable solution, but for this interval, we have broadly consistent wavevectors for the growing modes even though the velocities driving instabilities differ between the models.

\subsection{Statistical Overview}
\label{ssec:stats}

\begin{figure} 
 \centerline{\includegraphics[width=1.0\textwidth,clip=]{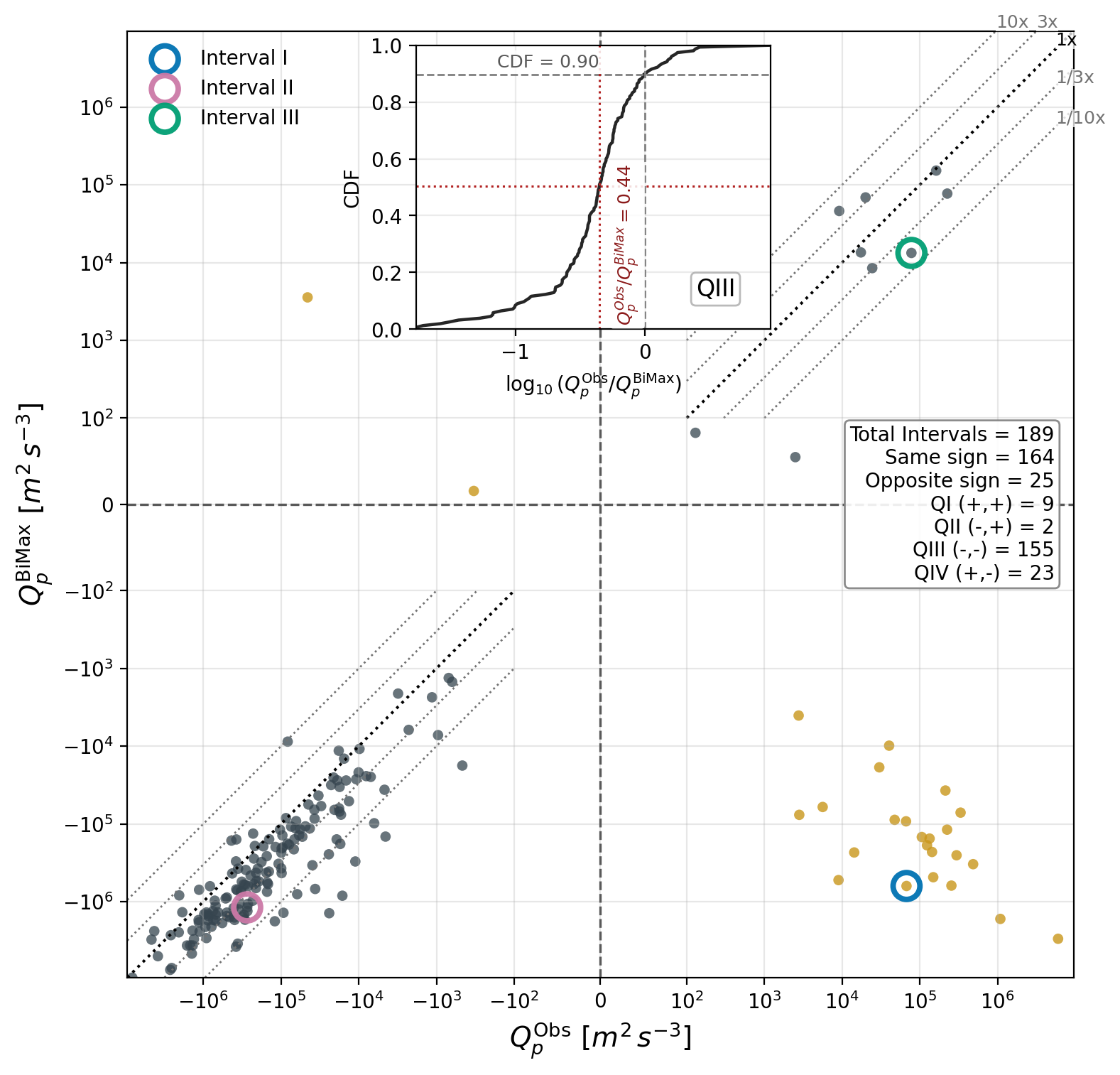}}
 \caption{
Comparison of integrated power emission and absorption, Eqn~\ref{eqn:Q}, for both the bi-Maxwellian and observed models.
The cumulative distribution function (CDF) for the 82\% of intervals for which both models predict net damping is shown as an inset.
The inset does not obscure any data points.
 }
 \label{fig:scatter}
 \end{figure}

Fig.~\ref{fig:scatter} shows a scatter plot of the evaluation of $Q_p^{\rm model}$, Eqn.~\ref{eqn:Q}, for the biMax model on the vertical and for the Obs model on the horizontal axis.
For the 342 SPAN-I intervals, 189 have coherent left-handed wave power over a finite range of frequencies.
Of those, 155(9) intervals have $Q_p<0(>0)$ for both models, and 23 have $Q_p^{\rm Obs}>0$ and $Q_p^{\rm BiMax}<0$.
The three intervals discussed in Sec.~\ref{ssec:intervals} are identified with circles.
\change{
The investigation of the coupling between the relative goodness of fit of the BiMax and Obs models and the resulting $Q_p$, not shown, does not reveal any significant correlations.
}

Quadrants I, II, and IV contain too few intervals for a robust statistical comparison.
The cases with both models with net damping have sufficient intervals for a statistical consideration.
By looking at the cumulative distribution function (CDF) of these points, we see that 90\% of the intervals have weaker damping in the Obs model than in the biMax model.
The median value of the ratio 
$Q_p^{\rm Obs}/Q^{\rm BiMax}$
is 0.44.

\section{Discussion}
\label{sec:discussion} 

Considering the linear solutions from our two models, the use of the observed VDF in \texttt{ALPS} produces more weakly damped solutions, and more net unstable intervals, than when using the best-fit two-component bi-Maxwellian model.
The observed VDF model produces unstable modes that occur more consistently \change{throughout the} interval than the biMax model, matching with the observed distribution of coherent left-handed power.

The physical origin of this difference is that the resonant interactions that drive emission and damping depend on velocity-space gradients and not on low-order velocity moments. 
A two-component bi-Maxwellian fit can reproduce the broad core-beam structure, but smooths over local features in the VDF near the resonant velocities that drive the observed waves. 
These features can introduce regions of emission that partially offset, or in some cases overcome, absorption elsewhere in the distribution.
As a result, moment-based stability estimates may miss growing solutions or systematically overestimate cyclotron damping when the measured VDF contains significant non-Maxwellian structure.

Future work in this avenue should consider the impact of using the observed $f_\alpha$ of the Helium population in the solar wind rather than a bi-Maxwellian representation, as $\alpha$ distributions have been shown to be important in driving waves in the inner heliosphere, in particular right-handed and/or oblique instabilities \citep{McManus:2024,Martinovic:2026}.
The impact on a broader set of wave modes, including right-handed fast magnetosonic waves associated with strongly drifting, high $T_\perp/T_\parallel$ beams, so-called \textit{hammerhead} distributions \citep{Verniero:2022,Das:2026a}, and more general directions of propagation should also be considered, as well as the impact of these non-Maxwellian VDFs on more complex processes proposed for the inner heliosphere, e.g.
the Helicity Barrier \citep{Meyrand:2021,Squire:2023}
or Cyclotron Breaking \citep{Yerger:2026}.

%
\begin{acks}
The SWEAP Investigation and this publication are supported by the PSP mission under NASA contract NNN06AA01C. 
The \texttt{ALPS} project received support from UCL's Advanced Research Computing Centre through the Open Source Software Sustainability Funding scheme. 
This study benefited from support by the International Space Science Institute (ISSI) in Bern, through ISSI International Team project 24‐612 (“Excitation and dissipation of kinetic‐scale fluctuations in space plasmas”) 
and 
the Nordita program "Synergies Between Astrophysical, Space, Laboratory, and Fusion Plasma Physics." 
DV is supported by STFC Consolidated Grant ST/W001004/1.
\end{acks}

%
%
%
%
\begin{dataavailability}
The spacecraft data for this Letter are openly available from the NASA Space Physics Data Facility \citep{Livi:2020:L2,Bale:2020}. 
Additional SWEAP data and information are available at the SWEAP web page
\url{http://sweap.cfa.harvard.edu/Data.html.}
\end{dataavailability}
\begin{codeavailability}
The ALPS code is available via an open source BSD 2‐Clause License at \url{https://github.com/danielver02/ALPS}
with a full tutorial on its use at \url{https://danielver02.github.io/ALPS/} \citep{ALPS:2023}.
\end{codeavailability}

\appendix

\begin{figure}
    \centering
    \includegraphics[width=1.0\linewidth]{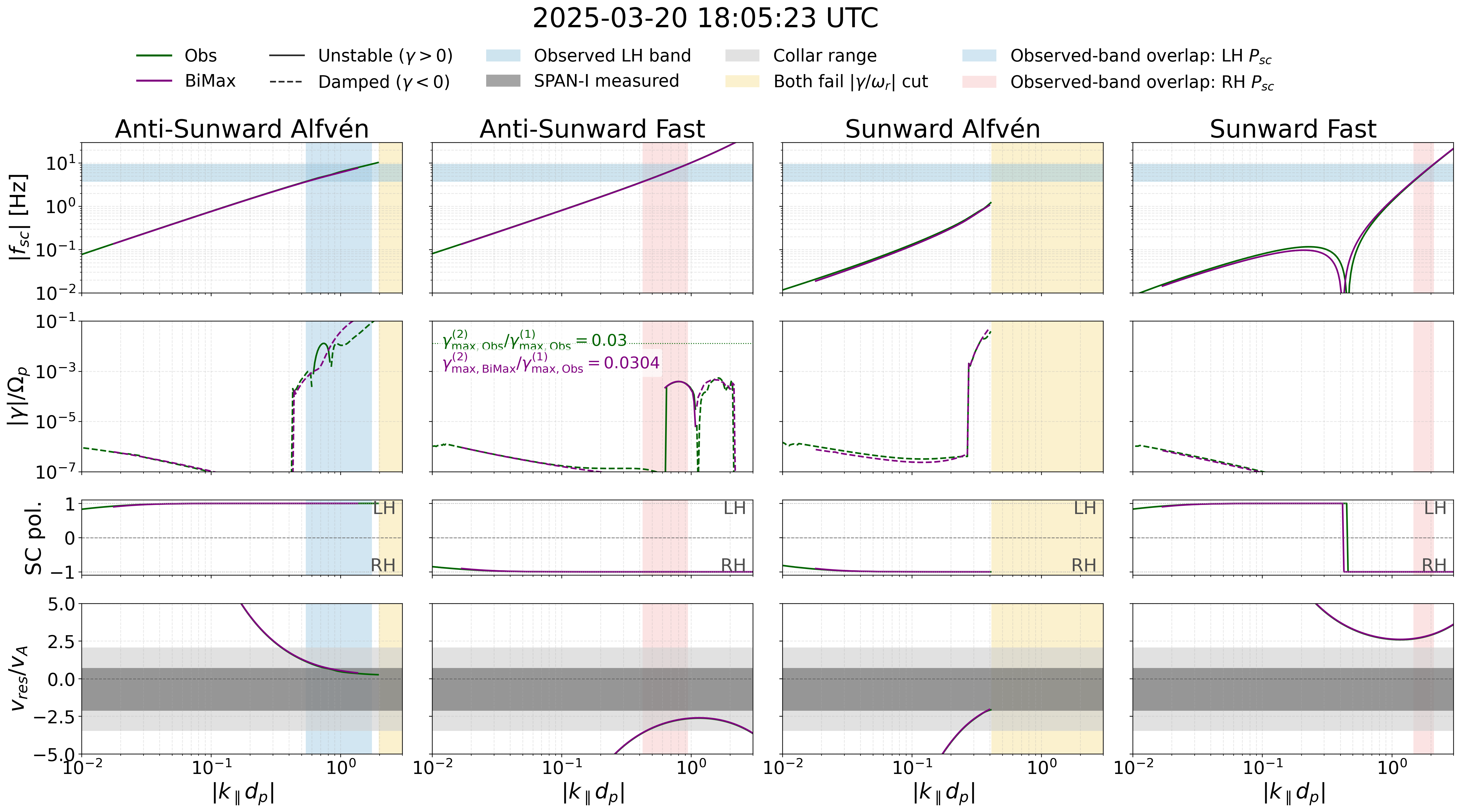}
    \caption{\change{Dispersion relation for four parallel propagating solutions, 
    Anti-Sunward \Alfven\ (first column), 
    Anti-Sunward fast (second), 
    Sunward \Alfven\ (third), and Sunward fast (fourth), for SPAN-I interval 2025-03-20: 18:05:23 for both the Obs (green) and BiMax (purple) models.
    The rows show as a function of parallel wavevector the magnitude of the spacecraft frame frequency (top row) compared to the observed frequency band of left-handed wave power (blue), 
    the damping (dashed lines) or growth (solid) rates (second row),
    the spacecraft frame polarization (third row), 
    and
    the resonant velocity $v_{\rm res}/v_A$ (fourth row), with bands indicating the SPAN-I resolved $v_\parallel/v_A$ (dark grey) as well as the collar width (light grey).
    Yellow regions indicate wavevectors for which $|\gamma/\omega_r|>0.3$, indicating a heavily damped solution.
    }}
    \label{fig:four-92}
\end{figure}

\begin{figure}
    \centering
    \includegraphics[width=1.0\linewidth]{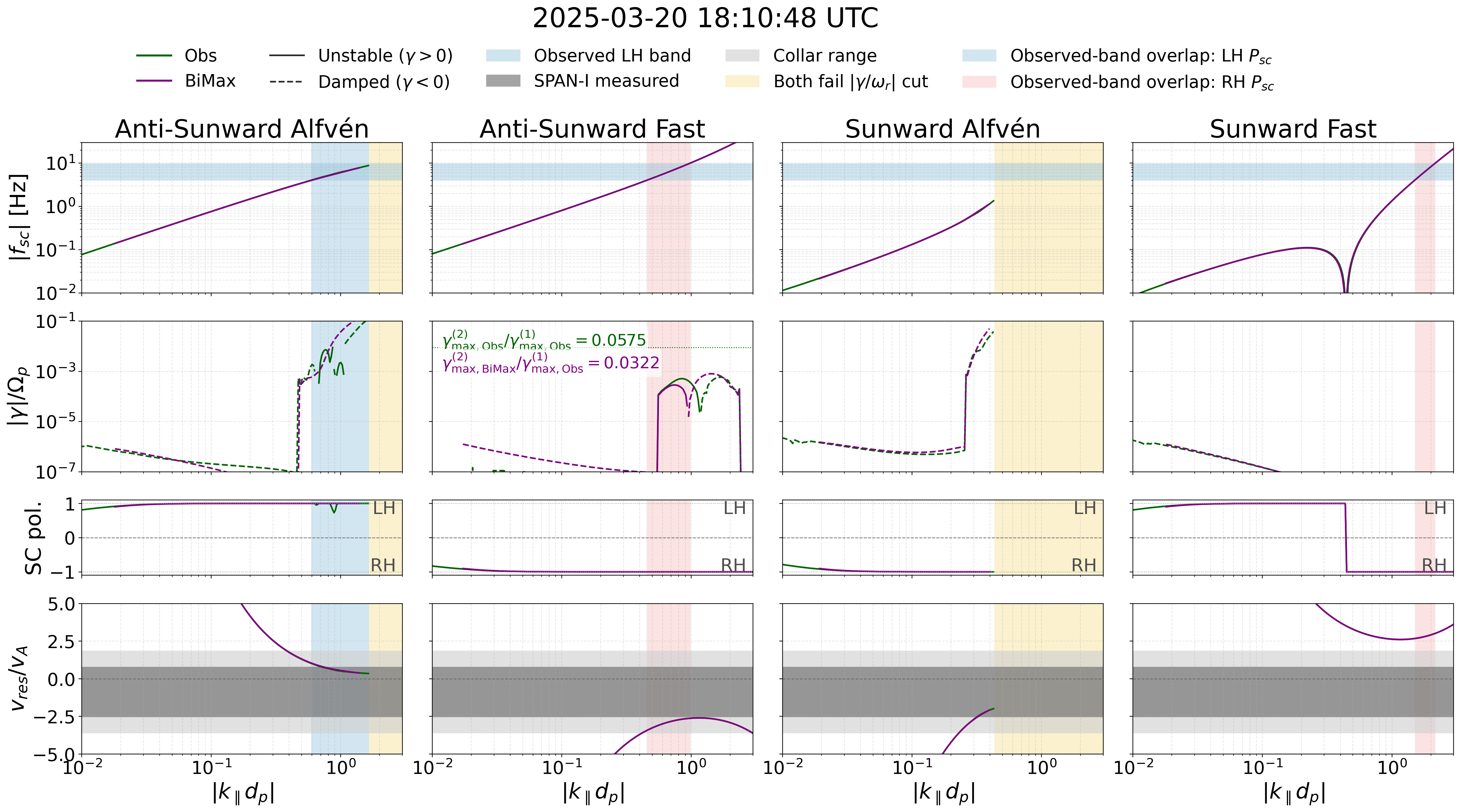}
    \caption{\change{Dispersion relations, in the same format as Fig.~\ref{fig:four-92}, for 2025-03-20:18:10:48.}}
    \label{fig:four-185}
\end{figure}

\begin{figure}
    \centering
    \includegraphics[width=1.0\linewidth]{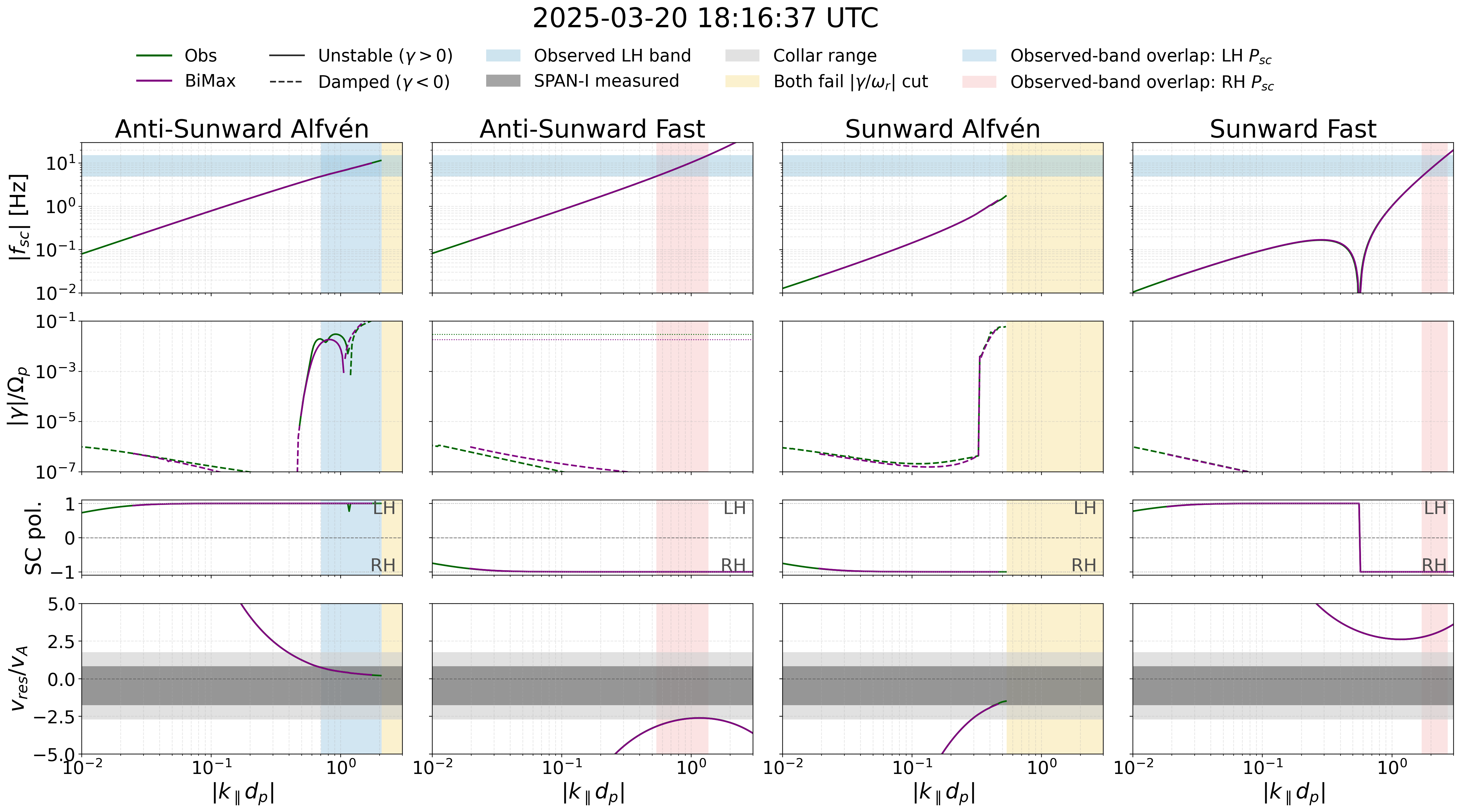}
    \caption{\change{Dispersion relations, in the same format as Fig.~\ref{fig:four-92}, for 2025-03-20:18:16:37.}}
    \label{fig:four-285}
\end{figure}

\change{To investigate  the role of the three other parallel propagating solutions, we investigate the full set of four parallel propagating waves, forward and backwards fast and \Alfven \ solutions, for the three times identified in Figs. \ref{fig:int1}, \ref{fig:int2}, and \ref{fig:int3}. 
These are shown in Figs.~\ref{fig:four-92}, ~\ref{fig:four-185}, and \ref{fig:four-285}.}

\change{For 18:05:23, Fig.~\ref{fig:four-92}, the Sunward \Alfven{}  wave becomes heavily damped prior to matching the observed $|f_{sc}|$ of the left-handed wave band.
Both the Sunward and Anti-Sunward fast modes have overlapping frequencies, but both have a right-handed spacecraft-frame polarization that is inconsistent with the observations.
The Anti-Sunward fast mode is resonant with the proton beam, driving a weak instability.
This unstable mode has a maximum growth rate of $3\%$ of the maximum growth rate of the Anti-Sunward \Alfven{} mode, further suggesting that it does not \change{play} a significant role in the dynamics of this interval.}

\change{
The interval around 18:10:48, Fig.~\ref{fig:four-185}, is similar to that around 18:05:23: the fast modes again fail to match the spacecraft-frame polarization, while the Sunward \Alfven{} mode becomes overly damped.
The Anti-Sunward fast mode is again unstable, but its maximum growth rate is $6\%$ of the maximum growth rate of the Anti-Sunward \Alfven{}  solution. 
Due to the lack of polarization agreement and orders of magnitude difference in growth rates, we rule out the fast mode as playing a significant role in this interval.}

\change{The interval around 18:16:37, Fig.~\ref{fig:four-285}, again shows support for only the Anti-Sunward \Alfven{} mode.
None of the other solutions have a polarization consistent with the observations, and the Anti-Sunward fast mode is stable.}

%

%
%
\bibliographystyle{spr-mp-sola}
\bibliography{master.bib}  
%
%
%
%

\end{document}